\documentclass[showpacs,aps,prl,twocolumn,superscriptaddress, letterpaper]{revtex4}
\pdfoutput=1
\usepackage{graphicx}
\usepackage{dcolumn}
\usepackage{bm}

\begin{document}

\title{Multi-Modal Properties and Dynamics of the Gradient Echo Quantum Memory}

\author{G.~H\'{e}tet}
\affiliation{ARC COE for Quantum-Atom Optics, Australian National
  University, Canberra, ACT 0200, Australia}

\author{J.~J.~Longdell}
\affiliation{Laser Physics Centre, RSPhysSE, Australian National
  University, Canberra, ACT 0200, Australia}
\affiliation{Department of Physics, University of Otago, Dunedin, New
Zealand}

\author{M. J. Sellars}
\affiliation{Laser Physics Centre, RSPhysSE, Australian National
  University, Canberra, ACT 0200, Australia}

\author{P. K. Lam}
\affiliation{ARC COE for Quantum-Atom Optics, Australian National
  University, Canberra, ACT 0200, Australia}

\author{B. C. Buchler}
\affiliation{Laser Physics Centre, RSPhysSE, Australian National
  University, Canberra, ACT 0200, Australia}

\begin{abstract}
We investigate the properties of a recently proposed Gradient Echo Memory (GEM) scheme for information mapping between optical and atomic systems.  We show that GEM can be described by the dynamic formation of polaritons in k-space. This picture highlights the flexibility and robustness with regards to the external control of the storage process. Our results also show that, as GEM is a {\it frequency-encoding} memory, it can accurately preserve the shape of signals that have large time-bandwidth products, even at moderate optical depths.  At higher optical depths, we show that GEM is a high fidelity multi-mode quantum memory.
\end{abstract}

\maketitle

The ephemeral nature of photons make them simultaneously useful and frustrating as messengers for quantum information. 
On the one hand, they travel with low absorption and are easy to produce and detect.  On the other hand, photons are hard to store in a manner that preserves quantum characteristics.  Over the last few years quantum information science has motivated the study of memories that can preserve the quantum characteristics of optical states \cite{liu}. Such quantum memories are key components of technologies such as single photon sources, that are required for many quantum information protocols; and quantum repeaters, that would allow propagation of quantum states over large distances \cite{Duan}.

There has been much work on light storage using techniques such as Electromagnetically Induced Transparency (EIT) \cite{liu} and Raman transfer \cite{Koz,Gor,Nunn}. Dynamic control of the reading and writing stages for single temporal modes can be used to optimise these systems and large classical efficiencies of 40\% have been reported for EIT in a vapour cell \cite{Novi08}.  Photon echo techniques are also candidates for light storage. They allow a high density of classical information to be stored \cite{Moss} with high efficiency \cite{Corn}.  Photon echo quantum memories have been proposed using controlled reversible inhomogeneous broadening (CRIB) \cite{mois01} and atomic frequency combs \cite{AFC}, both of which include $\pi$ pulses in the storage protocol. The Gradient Echo Memory (GEM) is a recently proposed variant of CRIB where the memory control is purely electro-optic and $\pi$-pulses are not required \cite{alex06,HetetPRL}. GEM is predicted to be 100\% efficient in the limit of large optical depths.  So far, experimental demonstrations of GEM have shown classical efficiencies of 15\% limited mostly by the optical thickness of the storage medium \cite{HetetPRL}. 

In this letter we show how GEM can be described using normal modes in $k$-space.  The analysis shows that the storage efficiency of GEM is not affected by the external control of the memory in time and demonstrates its potential to preserve the shape of pulses that have large time-bandwidth products (TBW) \cite{TBW}. Lastly, we show that GEM can simultaneously store a large number of temporal modes, allowing its implementation in recently proposed quantum repeater protocols \cite{sang06}.

\begin{figure}
  \centering
\includegraphics[width=\columnwidth]{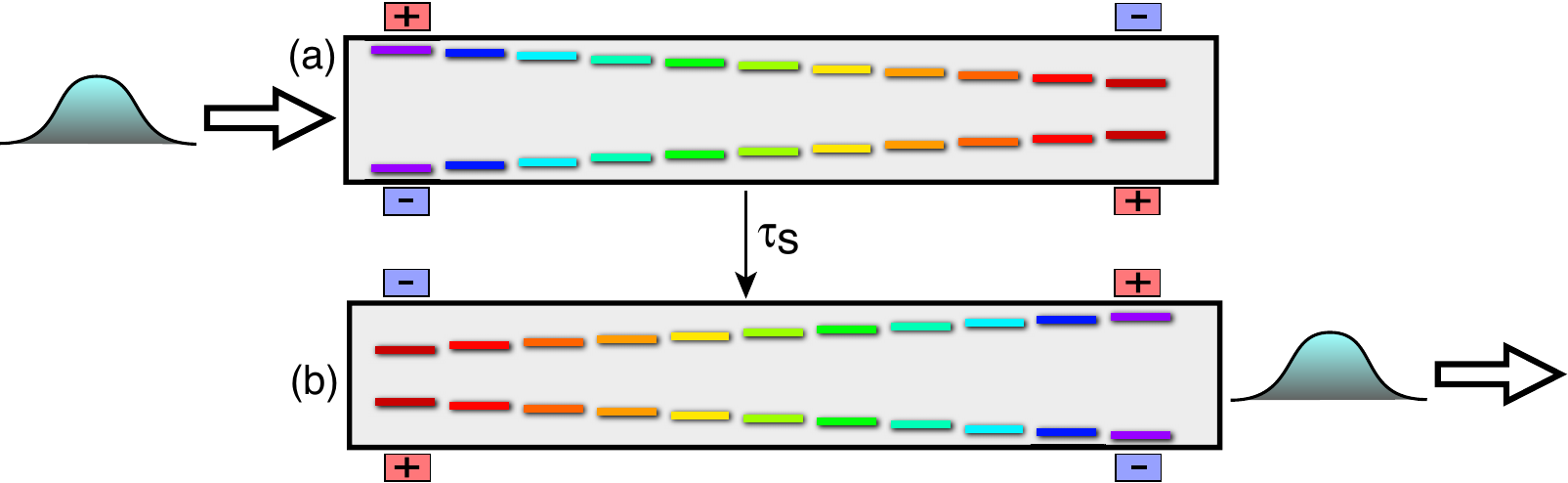}
  \caption{(Colour online)  (a) An ensemble of identical two-level atoms is Stark shifted by a linear electric field and the light field is sent into the storage medium. (b) After switching the polarity of the electric field, the input field comes out as a forward travelling echo.}
  \label{schematic}
\end{figure}

The GEM scheme is shown in Fig.~\ref{schematic}.  An ensemble of identical two-level atoms with homogeneous linewidth $\gamma$ is subjected to an electric field that varies linearly in $z$ causing a linearly varying Stark shift.  A light field enters the medium within a time interval $[t_1, t_2]$.  After some time $\tau_{s}/2$ the electric field gradient is flipped, leading to temporal reversal of the atomic phases.  At time $\tau_{s}$, the dipoles have all rephased and the input light emerges in the forward direction.  In the weak excitation limit, Heisenberg-Langevin equations can be solved by treating the optical field and atomic polarisation operators as c-numbers \cite{HetetPRL}. In a reference frame moving at the speed of light, the equations for the weak optical field, $\mathcal{E}$, and atomic polarisation, $\alpha$, are found to be \cite{HetetPRL}
\begin{eqnarray} \label{eq:mb1}
  \frac{\partial}{\partial t} \alpha(z,t) & = &  -\big[\gamma/2 + i \eta(t) z \big]  \alpha (z,t)  +ig \mathcal{E}(z,t) \nonumber\\
  \frac{\partial}{\partial z} \mathcal{E}(z,t) & = &i \mathcal{N}  \alpha(z,t),   \label{eq:mb2}
\end{eqnarray}
where $\mathcal{N}$ is the effective linear density of atoms and $g$ is the atomic coupling strength.  The linearly varying Stark shift is given by $\eta(t)z$, where the slope $\eta(t)$ can be controlled in time. We will neglect the decay rate $\gamma$ by assuming $\tau_{s} \ll 1/\gamma$.  The amount of light trapped in the medium depends on the optical depth. It has been shown \cite{JevonAnalytic} that the power in each frequency component $n$ within the bandwidth of the medium is stored with efficiency $\sqrt{\sigma_{n}}=1-e^{-2 \pi \beta}$ where $\beta=g\mathcal{N}/\eta$ is the optical depth.

\begin{figure}[!t]
  \centering
  \includegraphics[width=\columnwidth]{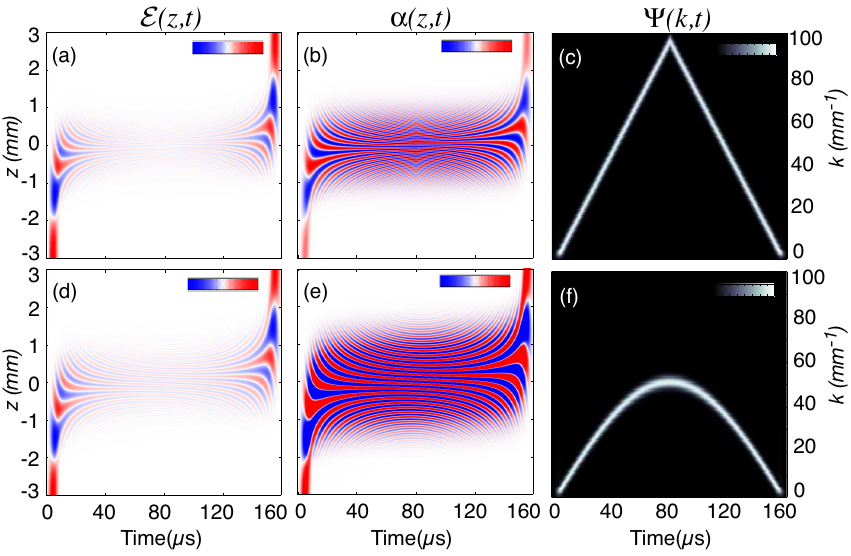}
  \caption{(Colour online) Top row: Abrupt switching of $\eta$, Bottom row: slow switching of $\eta$ using a $\tanh$ function with time constant $58\, \mu s$.  In all cases, the dipole ensemble spans $z=-3$ to $3\,$mm,  a Gaussian pulse enters the medium with its peak at $t=5\, \mu s$, the Stark shift is switched at $80\, \mu s$, and the optical depth $g\mathcal{N}/\eta = 3.3$.  }
  \label{gem1}
\end{figure} 

To describe the further evolution of the excitation {\it inside} the medium, we solve the problem within the time interval $t_2<t< \tau_s$. We make a plane wave decomposition of the optical and atomic fields via a spatial Fourier transform of Eqs.~(\ref{eq:mb1}), therefore introducing single mode operators in time and \emph{k-space}. In this $k$-space/time coordinate space, we can identify the normal modes $\Psi (k,t)=k \mathcal{E} (k,t)+\mathcal{N}  \alpha (k,t)$ which, from Eqs.~(\ref{eq:mb2}) have the following equation of motion
\begin{eqnarray}
\left( \frac{\partial}{\partial t} \right.-\left. \eta(t)
\frac{\partial}{\partial k}-\frac{ig\mathcal{N}}{k}\right)  \Psi (k,t)=0. \label{eqmotion}
\end{eqnarray}
From the Maxwell equation we see that the orthogonal normal modes $ \Phi (k,t)=k  \mathcal{E} (k,t)-\mathcal{N}  \alpha (k,t)$ are not excited regardless of the time dependence of $\eta(t)$.

Equation~(\ref{eqmotion}) shows that the rate at which $\Psi(k,t)$ evolves in the \emph{k-t} plane 
is controlled by the Stark shift $\eta(t)$.  This is because the  Stark gradient leads to a linear phase shift of $\alpha(z,t)$ as function of $z$, which is equivalent to a displacement in $k$-space.  The phase term in the propagation equation ($ig\mathcal{N}/k$) results from the interference between the atomic polarisation and light that is reradiated by the atoms.  As $\Psi(k,t)$  reaches high $k$ values, the amplitude of the electric field tends to zero and the phase shift becomes negligible.
Considering the quantum operators, we find the commutator $[\hat{\Psi},\hat{\Psi}^{\dagger}]=k^{2}+ \mathcal{N}^{2}$. We are, however, free to switch $\eta(t)$ to zero at any point in time so that the normal mode stays in the same spatial mode $k$.  In this case,  $\hat{\Psi}_{k}=\hat{\Psi}/\sqrt{k^{2}+\mathcal{N}^{2}}$ will be bosonic and can be identified as a polariton.  

Some numerical simulations are shown in Fig.~\ref{gem1} for an abrupt (top row) and gradual (bottom row) change in the sign of $\eta$. In both cases $\mathcal{E}(z,t)$ (left column) reaches a minimum magnitude at the switching point ($\tau_{s}/2$).  The phase evolution, best seen in $\alpha(z,t)$ (centre column) is clearly different, with the spatial oscillations in phase substantially faster for the case of abrupt switching.  This is manifest in the $k$-space evolution of the normal modes (right column).  For fast switching $\Psi(k,t)$ evolves at constant speed on the \emph{k-t} plane with a velocity reversal at $\tau_{s}/2$. For the slow switch $\Psi(k,t)$ slowly changes direction symmetrically about $\tau_{s}/2$. Regardless of the Stark-shift dynamics, $\Psi(k,t)$ propagates without loss. When $\Psi(k,t)$ returns to its input $k$ value the optical field leaves the storage medium.  The portion of light that is retrieved is $1-e^{-2\pi\beta}$ \cite{JevonAnalytic}, with the remainder of the light remaining trapped in the polaritonic mode. The total efficiency, including input and output losses, is thus $(1-e^{-2\pi\beta})^{2}$ for all spectral components within the Stark bandwidth.

\begin{figure}[!ht]
  \centering
  \includegraphics[width=\columnwidth]{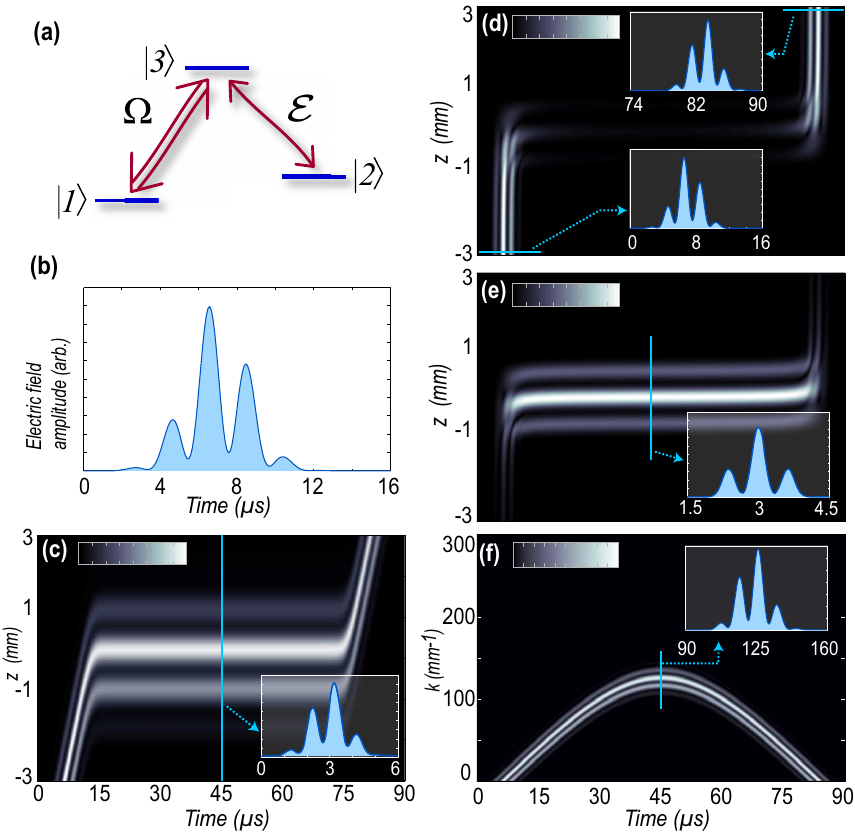}
  \caption{(Colour online) (a) The $\Lambda$ level scheme used for EIT. (b)~Modulated pulse used in the simulations.  (c)~EIT normal mode. Model parameters were $N$=5000, $g$=1 and $\Omega_c(0)=50$, normalized to the spontaneous emission rate. The control field was switched using $\tanh$ functions with a 2~$\mu$s time constant at 14 and 75~$\mu$s. In the case of GEM, (d)~shows $|\mathcal{E}|$, (e) shows $|\alpha|$ and (f) shows $|\Psi|$.  The switching was done at 45~$\mu s$ using a $\tanh$ function of time constant 20~$\mu s$. Optical depth $g\mathcal{N}/\eta = 3.3$. The efficiency in both schemes is close to 100\%.}
 \label{fig:pmodes}
\end{figure}

A coherent superposition of atomic polarisation and optical field also exists in EIT, in the form of dark state polaritons \cite{fleischhauer}. A qualitative comparison with the GEM normal mode is quite illuminating.  In EIT we have $N$ atoms in a $\Lambda$-level scheme as shown in Fig.~\ref{fig:pmodes}(a), where $\mathcal{E}$ is the light to be stored. $\Omega_c$ is the strong beam that is used to control the dynamics of the storage and thus plays an analogous role to $\eta$ in the GEM scheme.  The EIT normal mode will propagate slowly through the EIT medium while $\Omega_c$ is non-zero, and will become stationary when $\Omega_c=0$.  We consider here the case of the adiabatic switching of $\Omega_c$ discussed in \cite{fleischhauer}, which allows the storage of a frequency band within the EIT bandwidth and of temporal modes compressed within the atomic sample. 

To illustrate the properties of GEM and EIT we consider the storage of an amplitude modulated pulse, as shown in Fig.~\ref{fig:pmodes}(b). The EIT polariton is shown in Fig.~\ref{fig:pmodes}(c) and a spatial cross-section at 45~$\mu s$ shows that the temporal profile of the input pulse has been mapped into a spatial profile of the polariton.  Figure~\ref{fig:pmodes}(d) shows the absolute value of $\mathcal{E}(z,t)$ in a GEM system.  When the Stark shift is flipped, the echo emerges in the forward direction and, as demonstrated by the cross-sections, the output pulse is a time-reflected image of the input. This reversal could be corrected by using two cascaded GEM systems \cite{HetetPRL}. Figure~\ref{fig:pmodes}(e) shows the absolute value of $\alpha(z,t)$ in the GEM medium. The spatial cross-section at 45~$\mu s$ is the Fourier spectrum of the modulated input pulse.  This explicitly shows the frequency encoding nature of GEM. Figure~\ref{fig:pmodes}(f) again demonstrates this Fourier relationship by showing the evolution of the polaritons in $k$-space. Any cross-section in the k-axis shows the temporal profile of the pulse, as seen in the inset.

 A consequence of the above properties is that, provided the residual phase shifts and the decoherence $\gamma$ are negligible, the GEM storage efficiency can be maintained as the length of the signal is increased. For multi-mode light storage this means that, in the plane wave basis, the fidelity does not depend critically on the number modes stored. To demonstrate this result, let us introduce envelope mode functions defined in the region $[t_1,t_2]$,  as $u_{n}(t)=e^{i \omega_{n}t}/\sqrt{T}$, where $T=t_{2}-t_{1}$. This plane wave basis $\{u_{n}(t) \}$, is orthonormal and complete when $\omega_n=2\pi n/T$ where $n$ is an integer. In practice, a frequency cut-off will be given by the memory or detection bandwidth $\Delta\omega/2\pi$ so that the number of plane wave modes required to reconstruct any quantum state is $N_{mod}=T \Delta \omega/2\pi$. 
The multi-mode input field $\hat{\mathcal{E}}_{\rm in}(t)$ can be written as a unique linear superposition of those mode functions as $\hat{\mathcal{E}}_{\rm in}(t)=\sum_{n} \hat{a}^{\rm in}_{n} u_{n}(t)$ where $\hat{a}^{\rm in}_{n}$ is a single bosonic mode. We also write the mean total number of photons $\sum_{n}\langle(\hat{a}^{\rm in}_{n})^{\dagger}\hat{a}^{\rm in}_{n}\rangle=N_{ph}$. We then define the fidelity for the storage of $\hat{\mathcal{E}}_{\rm in}(t)$, as its correlation with the time-reversed output field $\hat{\mathcal{E}}_{\rm out}(t)$
\begin{eqnarray}
\mathcal{F}=\frac{1}{N_{ph}} \int_{t_1}^{t_2} dt \langle \hat{\mathcal{E}}^{\dagger}_{\rm out}(\tau-t)\hat{\mathcal{E}}_{\rm in}(t) \rangle,
\end{eqnarray}
where $\tau$ is the delay that maximizes $\mathcal{F}$. In general the output can be expressed as $\hat{\mathcal{E}}_{\rm out}(t)=\sum_{n} \sqrt{\sigma_{n}} \hat{a}^{in}_{n}u_n(-t)+\hat{b}(t)$, where $\hat{b}(t)$ includes the polaritonic modes left in the memory, and potential noise originating from the coupling with the other modes of $\hat{\mathcal{E}}_{\rm in}(t)$. We also define $\mathcal{F}^{r}=\mathcal{F}/\sqrt{\sigma}$ as a measure of pulse preservation independent of the total efficiency $\sigma= \sum_{n}\sigma_n$.

EIT based quantum memories can be optimised to store a single mode \cite{Gor} using temporal shaping of the control beam. Experiments have shown the technique to be highly effective \cite{Novi08}.  Optimal storage of multi-mode fields using EIT is, at this stage, still an open problem.  The adiabatic EIT protocol \cite{fleischhauer} modelled in Fig.~\ref{fig:pmodes} can store a quantum state comprised of multiple modes.  Compressing it into the storage medium requires a small group velocity for the whole signal, which in turn means a small global frequency bandwidth. The storage properties of GEM are rather different.  In a sense, GEM is a one-step realisation of the proposal made in \cite{scully}.  Many resonant systems are spread over a frequency interval to give a uniform time and frequency response.  In GEM, the input temporal modes are not spatially compressed within the medium so that all the frequency modes within the memory bandwidth are stored with equal efficiency, independent of optical depth.

\begin{figure}[!t]
  \centering
  \includegraphics[width=\columnwidth]{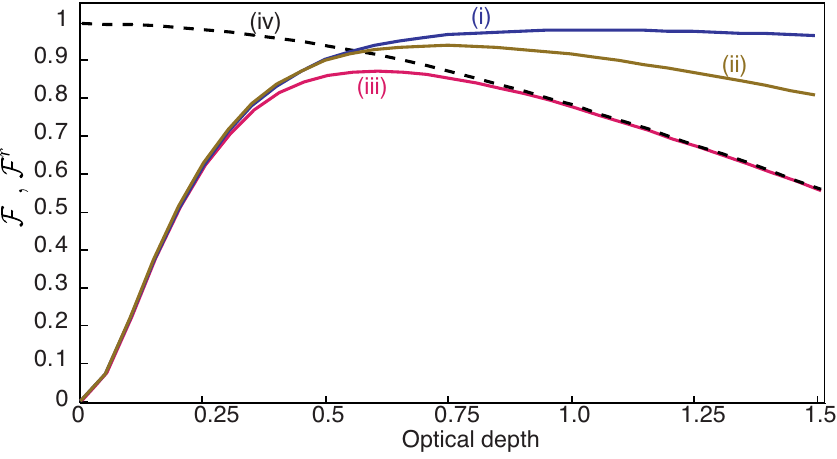}
  \caption{(Colour online) Multimode capacity of GEM in the plane wave basis with $\eta L/2\pi=8$MHz. Traces (i), (ii) and (iii) show the fidelity ($\mathcal{F}$) for all modes within the memory bandwidth and temporal extents $[35,45]\mu s$, $[20,60]\mu s$ and $[10,70]\mu s$ respectively.  Trace (iv) shows the mode shape preservation ($\mathcal{F}^{r}$) for modes within $[10,70]\mu s$.}
 \label{MM}
\end{figure}

To demonstrate the multi-mode capabilities of GEM, we calculate the storage fidelity for each of the plane-wave basis modes within the interval $T_1=[35,45]\mu s$. The simulations were performed using a memory bandwidth of $\Delta\omega/2\pi=\eta L/2\pi= 8$~MHz so that the number of modes that can be stored is $N_{mod}=T\Delta\omega/2\pi=80$.  All modes with a Fourier spectrum that lies within the memory bandwidth have identical storage fidelity.  Figure \ref{MM}~(i) shows the fidelity for all modes within the memory bandwidth and time interval $T_{1}$ when stored and retrieved independently.  When the optical depth is above 0.75, the fidelity is $>99$\% for all the modes. This shows that, at large optical depths, the noise term $\hat{b}(t)$ is negligible so that our memory only outputs linear combinations of these input modes. We conclude that any multimode field within this interval and bandwidth has the same fidelity $\mathcal{F}>99\%$.

We now evaluate the potential of GEM to store more modes by increasing the signal duration symmetrically around $t=40\mu s$. Figures (\ref{MM})~(ii) and (iii) show the result of simulations performed on plane wave modes within the memory bandwidth and time  intervals $[20,60]\mu s$ and $[10,70]\mu s$ respectively.  As the signal length increases, $\mathcal{F}$ degrades, but this is not irreparable.  Figure~\ref{MM}(iv) shows the mode overlap $\mathcal{F}^{r}$ corresponding to trace (iii).  This shows that at small optical depths, the pulse shape preservation is ideal, although the fidelity is poor due to low quantum efficiency. For high optical depths, on the other hand, the total efficiency $\sigma$ is ideal \cite{HetetPRL}, but the mode overlap is degraded due to the phase shifts in the storage system as the pulse length is now comparable to the storage time  \cite{HetetPRL,JevonAnalytic,MoiseevPRA}.  
In the present simulations, the effect is an identical and deterministic frequency shift $\delta$ to all the modes. We verified numerically that this shift can be repaired by applying an offset $\eta z-\delta$ to the Stark field during the read out stage.
After choosing the appropriate $\delta$ for each optical depth,  modes with longer temporal extents can be retrieved with the fidelity shown in Fig.~\ref{MM}(i). Another simple solution is to shift the output field frequency externally using an AOM. 

For even longer signals ($T>80\mu s$), the phase shifts yield slightly non-uniform frequency shifts \cite{JevonAnalytic,MoiseevPRA}. Increasing the storage time can then be used to improve the fidelity. A storage time of 330~$\mu$s for example, will keep the fidelity above 90\% in all cases shown in Fig.~\ref{MM} without extra-manipulations of the Stark-shifts. The excited state decay ($\gamma$) will eventually limit the length of signal, and number of modes, that GEM can store. One can then realise a quasi-two level atom using a Raman transition. The decoherence time will be given by the very slow ground state decay of a $\Lambda$ system \cite{OL}.

In conclusion, this study reveals several important features of the GEM scheme.  We have identified lossless polaritons that are shown to be insensitive to the switching dynamics of the Stark shift.  Even at low optical depths, this memory scheme preserves the pulse shape of signals that have large time-bandwidth product.  For larger optical depths, we have shown that GEM is a high fidelity multi-mode memory provided the storage time is shorter than the decoherence rate.

During the revision of the manuscript, we became aware of the work of \cite{nunn} that discusses the multi-mode properties of quantum memories.  The authors would like to thank Joseph Hope for useful discussions.  This work was supported by the Australian Research Council.


\begin{thebibliography}{99}
\bibitem{TBW}{TBW is defined as the product between the spectral full width at half maximum ($\Delta \omega$) and the duration $T$ of the signal}
\end{thebibliography}
\end{document}